\documentclass[twocolumn,aps,prl,preprintnumbers,superscriptaddress]{revtex4-2}
\usepackage[utf8]{inputenc}
\usepackage{amsmath}
\usepackage{amssymb}
\usepackage{graphicx}
\usepackage{textcomp}% Include figure files
\usepackage{bm}% bold math
\usepackage[right]{eurosym}
\usepackage{float}
\usepackage[english]{babel}
\usepackage{blindtext}
\usepackage{booktabs}
\usepackage{longtable}
\usepackage{xcolor}
\usepackage{tikz}

\begin{document}

\author{N. S. Blaj}
\affiliation{Institute of Physics, University of Kassel, 34132 Kassel, Germany}

\author{N. Scheel}
\affiliation{Department of Physics and Astronomy, Aarhus University, 8000 Aarhus C, Denmark}

\author{R. Rajni}
\affiliation{Institute of Physics, University of Kassel, 34132 Kassel, Germany}

\author{A. Ø. Lægdsmand}
\affiliation{Institute of Physics, University of Kassel, 34132 Kassel, Germany}

\author{J. Sadasivan}
\affiliation{Department of Physics,
Mahindra University Hyderabad, India}

\author{S.~De}
\affiliation{Sorbonne Université, CNRS, Laboratoire de Chimie Physique Matière et Rayonnement, UMR 7614, F-75005 Paris, France}

\author{S. R. Krishnan}
\affiliation{Department of Physics and QuCenDiEM-group, Indian Institute of Technology Madras, Chennai 600036, India}

\author{J. Bozek}
\affiliation{Synchrotron SOLEIL, St. Aubin, BP48, 91192 Gif sur Yvette Cedex, France}

\author{A. R. Milosavljevi\'c}
\affiliation{Synchrotron SOLEIL, St. Aubin, BP48, 91192 Gif sur Yvette Cedex, France}

\author{M. Mudrich}\email{mudrich@uni-kassel.de}
\affiliation{Institute of Physics, University of Kassel, 34132 Kassel, Germany}
\affiliation{Center for Interdisciplinary Nanostructure Science and Technology (CINSaT), University of Kassel, 34132 Kassel, Germany}

\title{X-ray photoelectron spectroscopy of Ar and Kr clusters formed in He nanodroplets}

\begin{abstract}
We report the first soft x-ray photoelectron spectroscopy (XPS) measurements of Ar and Kr clusters formed inside superfluid helium nanodroplets (HNDs) through consecutive pickup of dopant atoms. Ar and Kr atoms and clusters are selectively inner-shell ionized (Ar 2p, Kr 3d) using monochromatic soft x-ray synchrotron radiation. In the regime of strong doping, the electron spectra exhibit features characteristic of free Ar and Kr atoms as well as their clusters. The Kr cluster spectra agree well with literature data for bare Kr clusters. Backed by detailed simulations of the pickup process, this agreement indicates that the observed spectra originate from nearly bare Ar and Kr clusters, from which most or all He has evaporated in the course of cluster aggregation. These results establish HNDs as a platform for XPS of various types of molecular complexes and nanostructures.
\end{abstract}
\date{September 3, 2026}
\maketitle

\section{Introduction}
HNDs are widely used as nanometer-sized cryo-matrices for high-resolution spectroscopy of embedded molecules, molecular complexes, and nanostructures~\cite{slenczka2022molecules}. The main benefits of HNDs are that individual molecules are isolated in an ultracold (0.37 K), highly dissipative but weakly perturbing environment. Additionally, homogeneous and heterogeneous molecular complexes and even nanostructures can be formed by consecutive pickup of atoms or molecules of the same or different species by the HNDs~\cite{haberfehlner2015formation,messner2020shell, asmussen2023secondary}. 

High-resolution absorption and emission spectra of embedded molecules have been recorded in all spectral regions from microwave and infrared up to ultraviolet radiation. When the photon energy $h\nu$ matches or exceeds the absorption bands of HNDs, i.e. $h\nu\geq 21$~eV, Penning ionization of dopants through the transfer of energy from the excited He to the dopant, or charge transfer (CT) ionization through ionized He ($h\nu\geq 24.6$~eV), usually predominate over direct photoionization of dopants~\cite{laforge2024interatomic}. Although Penning ionization electron spectra (PIES) tend to be broad and lack structure, some useful information about the electronic structure of the embedded molecules and complexes can be inferred, e.g. outer-valence ionization energies~\cite{Buchta2013,BenLtaief2019,mandal2020penning,asmussen2023secondary,de2026valence} and the state of solvation by e.g. added water molecules~\cite{ltaief2026tracking}.

So far, direct photoionization of embedded atoms or molecules and the detection of emitted photoelectrons has mainly been achieved through multi-photon ionization using nanosecond lasers~\cite{radcliffe2004excited,loginov2005photoelectron,fechner2012photoionization,kazak2019photoelectron}.
Only two photoelectron spectroscopic studies have been reported in which dopants in HNDs were directly photoionized by one extreme-ultraviolet (EUV) photon~\cite{ben2020direct,ltaief2021photoelectron}. In those studies, dopants with exceptionally large photoabsorption cross sections compared with those of the HNDs were chosen (Xe at $h\nu\sim 100$~eV, coronene at $h\nu = 18.5$~eV) and the quality of the spectra was quite limited. 

On the other hand, heavier rare-gas clusters formed in free expansions have been extensively studied using synchrotron-based valence-level and core-level spectroscopies. A particular focus has been on the evolution of their electronic structure from atomic to condensed-phase, including site-specific core-level excitations, surface- and bulk-related binding-energy shifts, and local coordination effects~\cite{Knop1998,Tchaplyguine2003,Hatsui,Lundwall2006}.

The main challenge in photoionization experiments of molecules in HNDs is the low target density because the number density of HNDs is much lower than in typical molecular beam sources due to the much longer distance between nozzle and interaction region required to accommodate the doping chamber. Additionally, the proportion of dopant atoms to He atoms in doped HNDs usually amounts to $\lesssim0.1$\,\%; Consequently, if the EUV radiation contains even small portions ($\sim 1\,$\%) of higher harmonics, photoionization of HNDs and subsequent CT ionization of the dopants is likely to occur and mask the direct dopant photoionization signal. Furthermore, the residual gas (usually H$_2$O, N$_2$, O$_2$) in the interaction region is photoionized at any photon energy $h\nu\gtrsim 12$~eV and produces background signals that can far exceed the direct dopant photoionization signal depending on vacuum conditions. Therefore, to efficiently photoionize dopants in HNDs by EUV or x-ray radiation, conditions should be chosen such that i) the dopant concentration in the HND jet is high, ii) the photoabsorption cross section of the dopants is high relative to that of He and the residual gas, and iii) signals from photoionization of the dopants, e.g. photoelectrons, are detected selectively and sensitively. 

In the present experiment, we extend our previous EUV photoelectron spectroscopic studies~\cite{ben2020direct,ltaief2021photoelectron} to the soft x-ray region in view of exploring x-ray photoelectron spectroscopy (XPS) as a new approach to probing dopants in HNDs. We combine soft x-ray synchrotron radiation with electron spectroscopy using a hemispherical electron analyzer (HEA) to study rare-gas (Ar, Kr) clusters formed in HNDs as model systems. Distinct features appearing in the electron spectra are characteristic of Ar and Kr clusters formed in HNDs. These measurements were performed, however, in a high-doping regime in which nearly all He atoms had evaporated, leaving behind mainly bare Ar and Kr clusters. 
Nevertheless, this study paves the way for more extensive XPS studies benefiting from the unique properties of HNDs as weakly perturbing cryogenic matrices.

\section{Experimental setup}

\begin{figure}[t]
\centering
\includegraphics[width=\columnwidth]{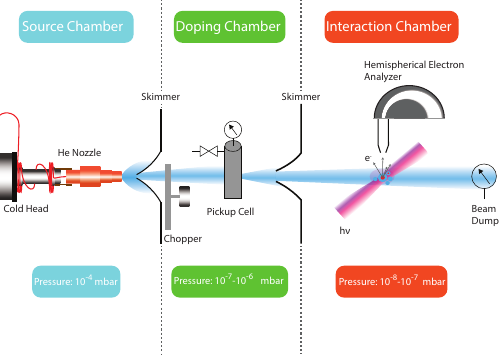}
\caption{Schematic of the HNDs beamline. HNDs are produced in the source chamber (SRC), skimmed, doped with Kr or Ar in a pickup cell, and collimated by a second skimmer before entering the interaction region, where photoionization occurs and electrons are analyzed with a hemispherical electron analyzer (HEA). The He flux in the transmitted HND jet is monitored by a pressure gauge mounted on a nipple downstream of the interaction region as a Pitot tube.}
\label{fig:setup_schematic}
\end{figure}

The experiments were performed using a HND beam apparatus coupled to the PLEIADES beamline at the synchrotron radiation facility SOLEIL near Paris. A schematic overview of the experimental setup is shown in Fig.~\ref{fig:setup_schematic}. The HND source has been described in earlier works~\cite{Scheidemann1993,Shcherbinin2018,Buchta2013,Mudrich2020,mandal2020penning,Buchta2013a,Lehmann99:639} and was used here in combination with a soft x-ray photoelectron spectroscopy endstation equipped with a HEA.

The setup comprised a source chamber, a doping chamber, and an interaction chamber. Downstream of the interaction region, the beam entered a beam dump equipped with a pressure gauge used to monitor the flux of the transmitted HND beam. 

HNDs were produced by continuous expansion of high-purity He gas through a cryogenic nozzle of diameter $5~\mu$m at a stagnation pressure of 50~bar. The nozzle temperature was varied between 13 and 17~K to tune the mean droplet size from $6.4\times10^{4}$ to $7\times10^{3}$ He atoms per droplet, respectively. Unless stated otherwise, the measurements discussed below were performed at $T = 16$~K, corresponding to an estimated mean initial droplet size of $\langle N_{\mathrm{He}}\rangle \approx 9700$.

The mean HND sizes were estimated using the HeNDS model developed by Raston~\cite{raston2021hends}, based on He equation-of-state calculations and the Knuth--Schilling--Toennies scaling approach. At $T=13$~K and a stagnation pressure of 50~bar, the expansion conditions lie close to the transition between the subcritical and supercritical expansion regimes, where the HeNDS model does not provide a reliable estimate of the mean HND size. 
Therefore, the corresponding value $\langle N_{\mathrm{He}}\rangle\approx6.4\times10^{4}$ was taken from Toennies and Vilesov~\cite{Toennies2004}. The comparison between the HeNDS calculations and the literature values is shown in Supplementary Fig.~S1.

Behind the nozzle, the central part of the HND beam was selected by a skimmer (0.4~mm diameter) and transmitted into the doping chamber, where the HNDs traversed a cylindrical pickup cell of length 18 mm. The pickup cell had entrance and exit apertures of 3~mm and 4~mm diameter, respectively.
The pickup cell was supplied with Kr or Ar through a gas-handling assembly, with the dopant pressure monitored by a capacitance gauge located on the gas inlet line.
Ar or Kr atoms were captured sequentially by the HNDs, resulting in the formation of dopant cluster cores inside the HNDs. At sufficiently high pickup levels, the \texttt{HeDopant} simulations discussed below indicate that nearly all He evaporates, leaving nearly bare dopant clusters.
The pickup conditions were adjusted by varying the gas pressure in the pickup cell. For Kr doping, the effective pressure in the pickup cell was varied in the range $(0.3{-}3.1)\times10^{-3}$~mbar. The background pressure due to He effusing from the source chamber into the doping chamber was taken into account. For Ar, representative spectra were recorded at a pickup pressure of $1.3\times10^{-3}$~mbar.

After the pickup cell, the doped HND beam passed through a second skimmer (3~mm aperture diameter) and entered the interaction chamber. The distance between the first skimmer and the interaction region was approximately 700~mm. In the interaction region, the HND beam was crossed with circularly polarized, monochromatized synchrotron radiation from the PLEIADES beamline. Photoelectrons were detected using a HEA (Scienta R4000). Because the analyzer sampled a finite solid angle in a fixed detection geometry, the measured relative intensities may be affected by photoelectron angular-distribution effects.

Photoelectron spectra were recorded in the Kr $3d$ and Ar $2p$ core-level regions mainly at photon energies $h\nu=130$ and 160~eV for Kr and at $h\nu=325$~eV for Ar. 
Additional Kr $3d$ measurements at other photon energies are summarized in Supplementary Fig.~S2.
The photon energies were chosen such that the photoionization cross sections of the dopants substantially exceed that of He. For example, at $h\nu=160$ eV, $\sigma(\mathrm{Kr}\,3d)\approx5.4$~Mbarn and $\sigma(\mathrm{He})\approx0.10$~Mbarn, whereas $\sigma(\mathrm{Ar}\,2p)\approx2.4$~Mbarn and $\sigma(\mathrm{He})\approx0.015$~Mbarn at $h\nu=325$~eV~\cite{YehLindau1985,Yeh1993}. 

The HEA settings were optimized for each measurement to balance spectral resolution and count rate. The monochromator exit slit was varied between 230 and 700~µm, corresponding to beamline energy resolutions ranging from approximately 0.2 to 0.4~eV. The overall energy resolution also included the contribution of the HEA and ranged from 0.29 to 0.54~eV depending on the measurement conditions. Further experimental parameters are summarized in the Supplementary Material.

Because the photoelectron signal associated with rare-gas species in HNDs was weak compared with the residual-gas photoionization background, foreground (FG) and background (BG) spectra were recorded sequentially under otherwise identical experimental conditions.
The HND beam reached the interaction region during the FG acquisition and was mechanically blocked using the beam chopper during the BG acquisition. Spectral decomposition and fitting were performed on the FG spectra, while the BG spectra were used to identify residual
gas-phase contributions.
For selected data sets, difference spectra were calculated as $I_{\mathrm{FG-BG}}=I_{\mathrm{FG}}-I_{\mathrm{BG}}$ to suppress the stationary residual-gas contribution and enhance spectral contributions associated with the doped HND beam.

The kinetic-energy scale of the HEA was calibrated using gas-phase Kr $3d$ spectra recorded during the beamtime. The reference line positions were calculated from the binding energies given in the literature~\cite{Vaughan1986}.
Small relative energy offsets, typically of the order of a few tens of meV, were observed in most of the sequential FG and BG acquisitions.
The presence of HNDs near the entrance of the electrostatic lens of the HEA appears to influence the electron energy measurement; however, the exact physical origin of the energy offsets could not be determined. 
Because all gas-phase atomic lines exhibited a displacement, although not always by exactly the same amount, the effect was attributed to a general experimental perturbation of the measured electron energies rather than to a dopant-specific binding-energy shift.
When a single relative offset could be reliably determined from the gas-phase atomic lines, the FG energy scale was aligned with the BG energy scale before subtraction. Otherwise, its possible influence on the difference spectrum was taken into account in the interpretation of the spectra.

The photon flux at the PLEIADES beamline depends on photon energy and beamline settings. Reference measurements performed with the HU80 undulator, the 600~lines/mm grating, and a monochromator exit slit of $100~\mu$m indicate fluxes on the order of $10^{13}$~photons/s around $h\nu=130$--160~eV and a few $10^{12}$~photons/s near $h\nu=325$~eV. Since different exit-slit settings were used in the present experiment, these values provide only an order-of-magnitude estimate rather than the absolute photon flux at the interaction region.

An order-of-magnitude estimate of the emitted photoelectron rate (see Supplementary Material), based on the experimental interaction geometry, the estimated HND density, the simulated average Kr pickup statistics, and the reference photon flux, gives a rate on the order of $10^{5}$ electrons/s before accounting for the limited acceptance angle and transmission of the HEA and finite detector efficiency. The maximum background-subtracted experimental count rate at the Kr (3d) peak was approximately $1.6\times10^{2}$ counts/s, about $10^{-3}$ of the estimated emitted photoelectron rate. Here, the experimental rate refers to the peak maximum. Consequently, spectra were accumulated over extended acquisition times to obtain sufficient counting statistics.

\section{Simulation of the doping process}

Interpreting the experimental spectra requires knowledge of both the number of dopant atoms captured by the HNDs and the corresponding droplet size after pickup. Under the strong-doping conditions employed in the present work, repeated pickup events deposit a substantial amount of energy into the HNDs, causing He evaporation and droplet shrinkage, thereby progressively reducing the pickup cross section. As a consequence, the pickup process deviates from the conventional Poisson model, which assumes a fixed HND size and a constant pickup cross section throughout the doping process.

To account for these effects, we employed the \texttt{HeDopant} simulation code developed by Sishodia and De~\cite{HeDopant}. The code has recently been applied to describe dopant pickup and cluster formation in HNDs~\cite{De2024}. Unlike analytical pickup models, \texttt{HeDopant} follows the coupled evolution of the droplet size, pickup probability, and He evaporation throughout the pickup process, providing size distributions of the dopant cluster and of the residual HNDs during and after the pickup process. Compared to the original implementation of \texttt{HeDopant}~\cite{HeDopant}, a few minor modifications were introduced to better represent the present experimental conditions. In particular, the treatment of the dopant thermal velocity and the description of dopant--dopant cohesive energies were refined. Details of these modifications are given in the Supplementary Material.

As shown in Supplementary Fig.~S3, the predicted mean Kr cluster size and the mean residual HND size depend only weakly on whether a fixed-size, lognormal, or exponential initial HND size distribution is assumed. However, the resulting Kr cluster-size distributions differ markedly in their peak positions, widths, and overall shapes.

\begin{figure}[t]
\centering
\includegraphics[width=\columnwidth]{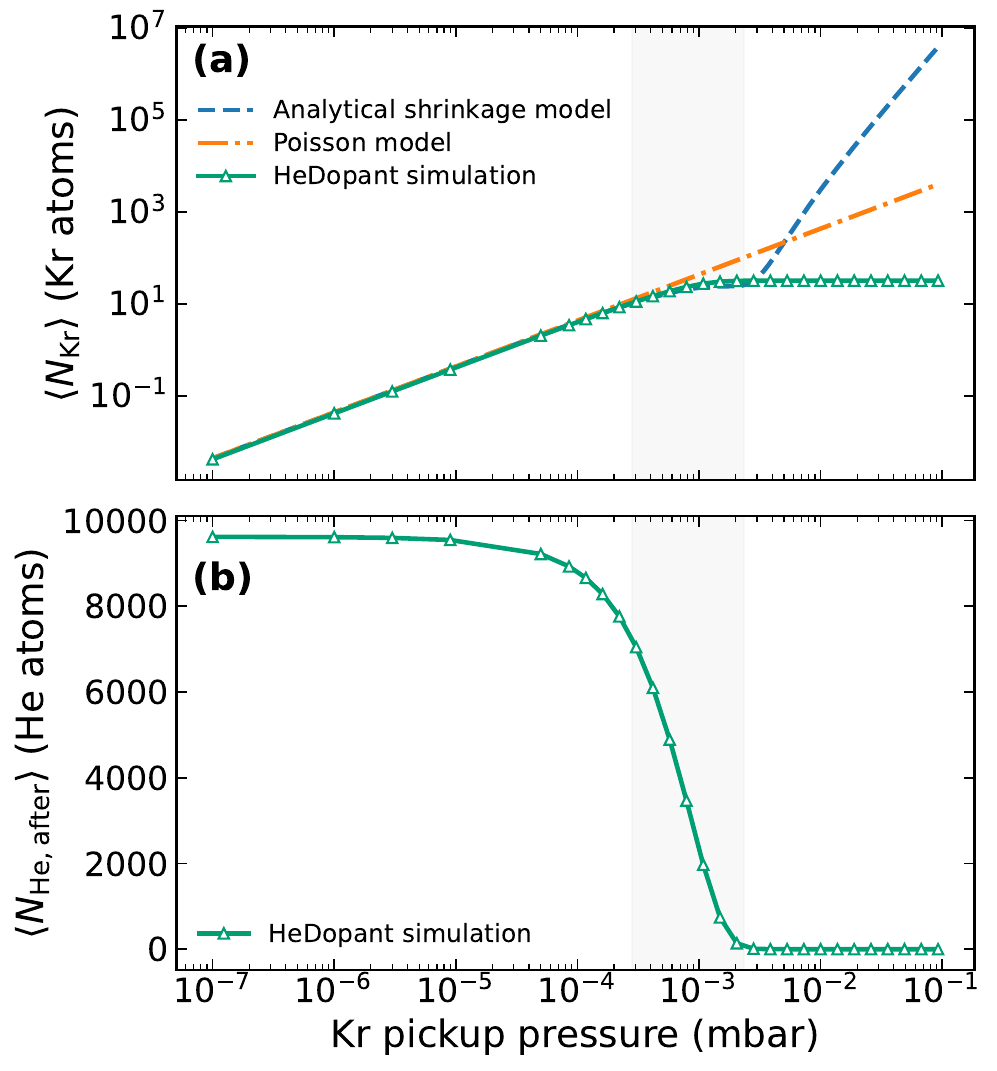}
\caption{Simulations of the pickup process of Kr atoms by HNDs as a function of the pickup pressure. (a) Mean number of captured Kr atoms per HND, $\langle N_{\mathrm{Kr}}\rangle$, compared with predictions by the conventional Poisson model and the analytical droplet-shrinkage model proposed in Ref.~\protect\cite{kuma2007laser}. The initial mean HND size was $\langle N_{\mathrm{He}}\rangle \approx 9700$. Beyond the complete-evaporation limit, formally continuing the droplet-shrinkage expression outside its physical range of validity produces the rapid upturn of the corresponding curve; this upturn does not represent further HND-mediated pickup.
(b) Mean HND size after pickup, $\langle N_{\mathrm{He,after}}\rangle$, due to progressive evaporative depletion of the droplets with increasing pickup pressure. The shaded region indicates the pressure range used in the current experiment.
}
\label{fig:Kr_pickup_models}
\end{figure}

Figure~\ref{fig:Kr_pickup_models} compares the \texttt{HeDopant} predictions with those of the conventional Poisson model and the analytical model proposed by Kuma~\textit{et al.}, which accounts for droplet shrinkage during the doping process~\cite{kuma2007laser}. At low pickup pressures, all three approaches predict essentially identical mean Kr pickup numbers. Above approximately $10^{-4}$~mbar, the conventional Poisson model deviates noticeably because it neglects the progressive reduction of the HND size and pickup cross section. In contrast, the droplet-shrinkage model remains in close agreement with the \texttt{HeDopant} simulations up to approximately $2\times10^{-3}$~mbar, where nearly all He atoms have evaporated from the droplets. Beyond this point, the analytical expression is outside its physical range of validity; the rapid upturn shown in Fig.~\ref{fig:Kr_pickup_models} is therefore a nonphysical extrapolation rather than a prediction of continued pickup.
The agreement up to the complete-evaporation limit confirms that the droplet-shrinkage model provides a reliable description of the mean pickup number over most of the experimentally relevant pickup-pressure range, including the strong-doping regime.
However, determining the dopant cluster-size distributions and the number of evaporated He atoms requires a more detailed treatment as provided by \texttt{HeDopant}. In the following analysis, we use the mean Kr cluster sizes and residual HND sizes predicted by \texttt{HeDopant}.

For the strong-doping conditions employed in the present experiment, these results indicate that the detected Kr clusters are contained in strongly depleted HNDs and are nearly stripped of He at the highest pickup pressures. Consequently, the characteristic dopant-related XPS features, including the observed cluster--gas energy shifts, are expected to reflect primarily Kr cluster formation and the electronic structure of the aggregates rather than the influence of a substantial surrounding He shell. However, a small contribution from the residual He environment cannot be ruled out.

\section{Results and discussion}

We first present representative Kr $3d$ and Ar $2p$ photoelectron spectra. We then focus on a systematic study of the Kr $3d$ spectra as a function of pickup pressure, initial HND size and photon energy, and we compare the extracted core-level shifts with previous measurements on free rare-gas clusters. Auger electron spectroscopy is additionally used to obtain complementary information about Kr aggregation under strong-doping conditions.

\subsection{Core-level photoelectron spectra of Kr and Ar clusters formed in HNDs}

\begin{figure}[t]
\includegraphics[width=\columnwidth]{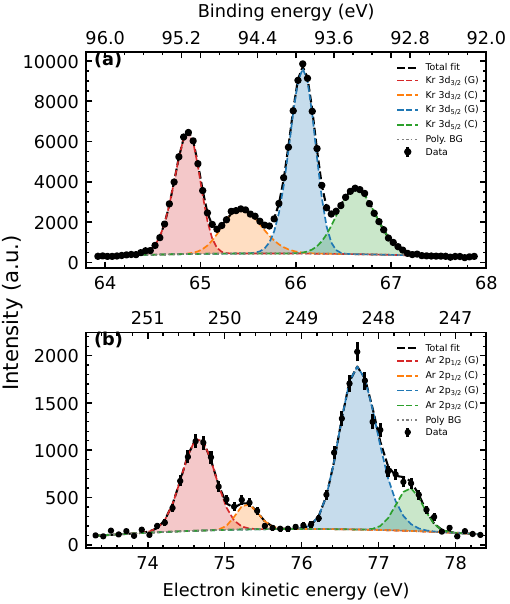}
\caption{
Representative soft x-ray photoelectron spectra of Kr and Ar dopants in HNDs produced at a nozzle temperature of $T = 16$~K, corresponding to an estimated initial mean HND size of $\langle N_{\mathrm{He}}\rangle \approx 9700$.
(a) Kr $3d$ spectrum recorded at $h\nu = 160$~eV under strong-doping conditions at a Kr pickup pressure $\approx1.3\times10^{-3}$~mbar.
(b) Ar $2p$ spectrum recorded at $h\nu = 325$~eV at an Ar pickup pressure $\approx1.3\times10^{-3}$~mbar.
\texttt{HeDopant} simulations yield average pickup numbers of $\langle N_{\mathrm{Kr}}\rangle\approx29$ and $\langle N_{\mathrm{Ar}}\rangle\approx37$, with corresponding
mean residual HND sizes after pickup and He evaporation of the order of $10^{3}$ and $6\times10^{2}$ He atoms, respectively.
The lower axes show the electron kinetic energy, while the upper axes display the corresponding binding energy.
Black dots represent the experimental data and solid red lines the total fit curves. Individual gas-phase and cluster-related fit components are shown as dashed lines together with the polynomial background contribution.
}
\label{fig:Kr_Ar}
\end{figure}

To identify the cluster-related spectral components and core-level energy shifts associated with rare-gas cluster formation in HNDs, we examined representative soft x-ray photoelectron spectra of Kr- and Ar-doped HNDs. Figure~\ref{fig:Kr_Ar} shows spectra recorded under strong-doping conditions without background subtraction.
The HNDs were produced at a nozzle temperature of $T=16$~K, corresponding to an estimated mean initial HND size of $\langle N_{\mathrm{He}}\rangle\approx9700$. The Kr spectrum was measured in the $3d$ region at $h\nu=160$~eV, whereas the Ar spectrum was recorded in the $2p$ region at $h\nu=325$~eV. Both measurements were performed at pickup pressures of $1.3\times10^{-3}$~mbar.  
According to the \texttt{HeDopant} simulation, these conditions correspond to average pickup numbers of $\langle N_{\mathrm{Kr}}\rangle\approx29$ and $\langle N_{\mathrm{Ar}}\rangle\approx37$. The corresponding mean residual HND sizes after pickup and evaporative cooling are approximately 1200 and 610 He atoms for the Kr and Ar measurements, respectively.

Both spectra contain narrow atomic features, predominantly assigned to gas-phase Ar and Kr atoms present in the interaction region, and broader features shifted toward lower binding energies, attributed to clusters formed inside the HNDs. In Fig.~\ref{fig:Kr_Ar}(a), the gas-phase Kr $3d_{3/2}$ and $3d_{5/2}$ components are shown in red and blue, respectively, while the corresponding cluster-related components are shown in orange and green. The same color scheme is used in Fig.~\ref{fig:Kr_Ar}(b) for the Ar $2p_{1/2}$ and $2p_{3/2}$ components.
A characteristic feature of both spectra is the shift of the cluster-related peaks toward lower binding energies relative to the gas-phase atomic lines, amounting to approximately $0.6$~eV for both Kr and Ar.
Such shifts are characteristic of condensed rare-gas systems and are generally attributed to final-state polarization screening, whereby the surrounding atoms partially screen the photohole created by photoionization~\cite{Hatsui,Tchaplyguine2003,Lundwall2006}.

To quantify the observed spectral features, the photoelectron spectra were analyzed using a nonlinear least-squares fitting procedure.
Each spectrum was modeled as a sum of skew-normal line shapes~\cite{Azzalini1985} representing the individual gas-phase and cluster-related contributions, together with a quadratic polynomial that empirically models the background,
\begin{equation}
I(E) = \sum_i S_i(E) + (aE^2 + bE + c).
\end{equation}
Here, $S_i(E)$ denotes the skew-normal profile of component $i$. 
For convenience, the skew-normal profiles were parametrized directly in terms of the integrated area $Q$, the mean kinetic energy $\bar{E}$, the standard deviation $s$, and the skewness parameter $\alpha$. For $\alpha\neq0$, the conventional location and scale parameters do not coincide with the mean and standard deviation of the distribution.
This choice leaves the line shape unchanged but provides direct access to the quantities used in the quantitative analysis and facilitates the application of physically motivated constraints. An equivalent fit could be performed using the conventional parameters and transforming the fitted values afterward. Accordingly, the peak positions reported below correspond to the fitted mean kinetic energies $\bar{E}$.
The fits were performed using weighted least squares with Poisson counting uncertainties, $\sigma_i=\sqrt{N_i}$. For selected spectra, physically motivated constraints were applied only where necessary to stabilize weak or strongly overlapping components, for example by fixing spin-orbit separations or relative intensities between corresponding components. Further details of the skew-normal parametrization and the implementation of the fitting constraints are provided in the Supplementary Material.

The reported $1\sigma$ statistical uncertainties were obtained from the fit covariance matrix and, where constraints were applied, propagated through the imposed parameter relations; they do not include systematic effects associated with the finite instrumental resolution or residual-gas background. 

From the fit of the Kr spectrum recorded at $h\nu = 160$~eV we obtain the mean kinetic energies of the gas-phase $3d_{3/2}$ and $3d_{5/2}$ components $64.85$~eV and $66.06$~eV, respectively. The corresponding cluster-related components appear at $65.45$~eV and $66.64$~eV, yielding gas-to-cluster kinetic-energy shifts of about $0.60$~eV and $0.58$~eV, respectively. The cluster-related peaks are broader than the gas-phase contributions, with fitted standard deviations increasing from about $0.14$~eV for the atomic lines to approximately $0.20$--$0.21$~eV for the cluster peaks. A similar fit of the Ar spectrum results in mean kinetic energies of the gas-phase $2p_{1/2}$ and $2p_{3/2}$ components $74.66$~eV and $76.76$~eV, respectively, while the cluster-related contributions appear at about $75.30$~eV and $77.40$~eV. Thus, the gas-to-cluster kinetic-energy shifts amount to approximately $0.64$~eV for both spin--orbit components.

The asymmetric shape of the fitted Kr cluster features likely reflects an unresolved superposition of contributions from Kr atoms in different coordination environments. In published spectra of pure Kr clusters, the Kr $3d$ signal was decomposed into dimer, corner, edge, and face/bulk contributions~\cite{Hatsui}. Because the dimer component was reported to be nearly unshifted from the atomic line, whereas the corner and edge components were observed at lower binding energies, their unresolved superposition may contribute to the asymmetric line shapes in the present, less-resolved spectra.

\begin{figure}[t]
\centering
\includegraphics[width=\columnwidth]{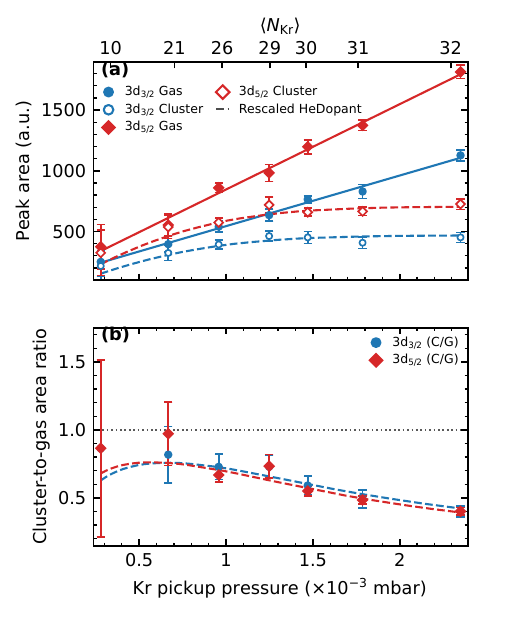}
\caption{
(a) Photoelectron peak areas of the Kr $3d_{3/2}$ and $3d_{5/2}$ lines measured at $h\nu=130$~eV for gas-phase atoms and Kr clusters embedded in HNDs as a function of the Kr pickup pressure. The top axis indicates the mean number of dopant atoms per HND, $\langle N_{\mathrm{Kr}}\rangle$, obtained from the \texttt{HeDopant} simulations.
The solid blue and red lines represent linear fits to the gas-phase data. The dashed blue and red curves represent the pressure dependence of $\langle N_{\mathrm{Kr}}\rangle$ predicted by \texttt{HeDopant}, independently rescaled in amplitude to fit the two cluster-related peak-area series using multiplicative factors determined by weighted
least-squares fits; they are included as guides to the eye. (b) Cluster-to-gas peak-area ratios for the $3d_{3/2}$ and $3d_{5/2}$ components.  The dashed curves were obtained by dividing the corresponding rescaled \texttt{HeDopant} curves in panel (a) by the linear fits to the gas-phase peak areas; no additional fit was applied to the ratio data. The horizontal dotted line indicates unity. Error bars denote $1\sigma$ fitting uncertainties.
}
\label{fig:Kr_area_ratio}
\end{figure}

\begin{figure}[t]
\centering
\includegraphics[width=\columnwidth]{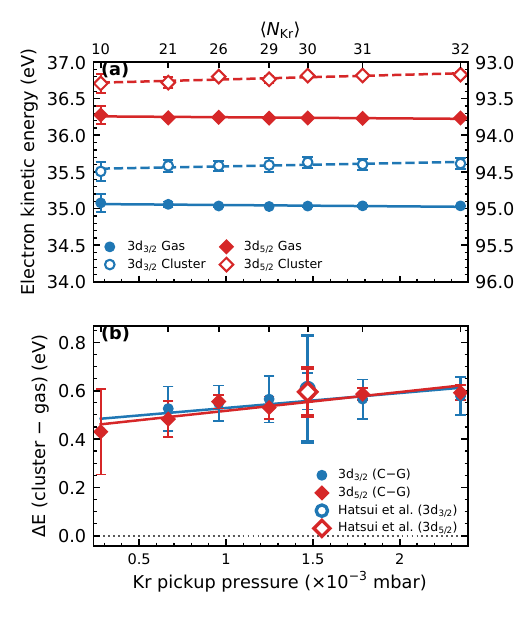}
\caption{
(a) Electron energies $E_e$ of the Kr $3d_{3/2}$ and $3d_{5/2}$ photoelectron lines measured at $h\nu=130$~eV for gas-phase atoms and Kr clusters embedded in HNDs as a function of the Kr pickup pressure. The right axis shows the corresponding binding energies $E_B$, and the top axis indicates the mean dopant number per HND $\langle N_{\mathrm{Kr}}\rangle$.
(b) Gas-to-cluster kinetic-energy shifts $\Delta E$ for the $3d_{3/2}$ and $3d_{5/2}$ components. Open symbols denote digitized literature values for pure Kr clusters of comparable mean size~\protect\cite{Hatsui}. Solid and dashed lines represent linear fits to the gas-phase and cluster data, respectively; error bars denote 1$\sigma$ fit uncertainties.
}
\label{fig:Kr_shift_vs_doping}
\end{figure}

\subsection{Dependences on pickup pressure, droplet size, and photon energy}

To investigate how the intensities and energy shifts of the cluster-related Kr $3d$ components evolve with increasing pickup pressure, Kr $3d$ photoelectron spectra were recorded while keeping the remaining experimental conditions constant.
Figures~\ref{fig:Kr_area_ratio} and \ref{fig:Kr_shift_vs_doping} summarize the corresponding measurements recorded at $h\nu = 130$~eV. The top axes indicate the mean number of picked up Kr atoms per HND, $\langle N_{\mathrm{Kr}}\rangle$, obtained from the \texttt{HeDopant} simulations. Figure~\ref{fig:Kr_area_ratio}(a) shows the integrated peak areas extracted from the spectral fits for the gas-phase and cluster-related Kr $3d_{3/2}$ and $3d_{5/2}$ components as a function of pickup pressure. For comparison, the pressure dependence of
$\langle N_{\mathrm{Kr}}\rangle$ predicted by \texttt{HeDopant}
was independently rescaled in amplitude to fit the experimental values for the cluster-related peak-area. The resulting dashed curves match the experimental data well, which confirms the validity of the pickup simulation in the studied range of pickup pressures.

At low pickup pressures, the spectra contain comparable gas-phase and cluster-related contributions. With increasing pickup pressure, both contributions increase, but the gas-phase signal grows more rapidly. The cluster-to-gas intensity ratio shown in Fig.~\ref{fig:Kr_area_ratio}(b) quantifies the relative prominence of the HND-associated cluster signal with respect to the effusive gas-phase Kr background. Its decrease from values close to unity at low pickup pressures to approximately $0.4$ at the highest investigated pressures shows that the gas-phase background increasingly dominates the measured spectra.

We attribute this behavior to the different pressure dependences of the two contributions. As the pressure in the pickup cell is raised, more free Kr atoms reach the interaction region and are directly photoionized. At the same time, repeated pickup and the associated He evaporation shrink the HNDs and progressively reduce their pickup cross section, causing the cluster-related signal to increase more slowly and eventually tend to level off.

The pickup pressure therefore serves as the experimental control parameter for cluster growth, but, in the strong-doping regime, the mean Kr cluster size does not increase proportionally with pressure. The nonlinear relationship between these quantities is obtained from the \texttt{HeDopant} simulations shown in Fig.~\ref{fig:Kr_pickup_models}, which account for pickup-induced He evaporation and the resulting changes in droplet size and pickup cross section.

Starting from an initial mean HND size of $\langle N_{\mathrm{He}}\rangle \approx 9700$, the mean number of remaining He atoms decreases from several thousand at the lowest investigated pressures to only a few hundred atoms or less at the highest pressures, while the mean number of captured Kr atoms approaches $\langle N_{\mathrm{Kr}}\rangle \approx 32$. Under these conditions, the measured spectra likely originate from compact Kr clusters embedded in strongly depleted HNDs and, at the highest doping level, from nearly He-free clusters.

An important question is how the electronic structure of the Kr clusters evolves with their growing size. The evolution of the peak positions as a function of pickup pressure obtained from the fits is shown in Fig.~\ref{fig:Kr_shift_vs_doping}(a). Both the gas-phase and cluster-related components exhibit nearly constant electron energies over the investigated pressure range. Thus, the extracted gas-to-cluster kinetic-energy shifts shown in Fig.~\ref{fig:Kr_shift_vs_doping}(b) remain approximately constant within the experimental uncertainties, with values $\Delta E \approx 0.5$--$0.6$~eV for both spin--orbit components. Despite the pronounced increase in the mean Kr cluster size predicted by the \texttt{HeDopant} simulations, the gas-to-cluster kinetic-energy shift changes only weakly with pickup pressure. This indicates that the local electronic environment responsible for core-hole screening becomes cluster-like already at an early stage of aggregation and is only weakly affected by subsequent cluster growth over the investigated size range.

On the binding-energy scale, the cluster-related components are shifted toward lower energy by approximately $0.60$~eV for the Kr $3d_{3/2}$ component and $0.58$~eV for the Kr $3d_{5/2}$ component. These values agree well with previous measurements of pure Kr clusters formed in a free expansion~\cite{Hatsui}, shown as open symbols in Fig.~\ref{fig:Kr_shift_vs_doping}(b). For clusters with $\langle N_{\mathrm{Kr}}\rangle=30$, close to the mean size predicted under the present conditions, corner- and edge-site shifts of $0.491$ and $0.645$~eV were reported with relative intensities of $18\%$ and $24\%$, respectively~\cite{Hatsui}. Their intensity-weighted mean, $0.579$~eV, closely matches the cluster shifts measured here. This agreement is consistent with the fitted cluster feature containing unresolved corner- and edge-site contributions rather than representing a single site-specific component.

\begin{figure}[t]
\centering
\includegraphics[width=\columnwidth]{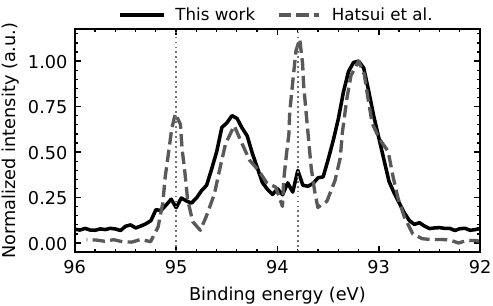}
\caption{
Comparison of the present FG$-$BG Kr $3d$ difference spectrum measured at $h\nu=160$~eV with a digitized spectrum of free Kr clusters from Ref.~\protect\cite{Hatsui}.
Before calculating the difference spectrum, the FG kinetic-energy scale was shifted by $0.042$~eV toward lower kinetic energy relative to BG to compensate for the relative FG--BG energy offset.
To compare the resulting FG$-$BG spectrum with the digitized literature spectrum, their remaining relative energy offset was corrected by applying an additional rigid shift $\Delta E$, determined from the mean peak positions obtained using two-Gaussian fits.
Both spectra are normalized to the maximum of the low-binding-energy feature in the range $93.0{-}93.5$~eV.
The vertical dotted lines indicate the atomic reference energies.
For the literature spectrum, a variable fraction of the atomic Kr contribution was subtracted from the cluster-beam data.
}
\label{fig:hatsui_comp}
\end{figure}

A more direct comparison is presented in Fig.~\ref{fig:hatsui_comp}, which compares the present
FG$-$BG Kr $3d$ difference spectrum measured at $h\nu=160$~eV with a digitized spectrum of free Kr clusters
reported previously~\cite{Hatsui}. For the literature spectrum, a variable fraction of the atomic Kr contribution was
subtracted from the cluster-beam data. Thus, both spectra emphasize cluster-related contributions, although the
subtraction procedures are not identical. To compensate for the HND-beam-associated energy offset described above, the FG kinetic-energy scale used for Fig.~\ref{fig:hatsui_comp} was shifted by $0.042$~eV toward lower kinetic energy relative to BG before subtraction. This corresponds to a shift of the same magnitude toward higher binding energy.

After normalization to the maximum of the low-binding-energy feature near $93$~eV, the overall cluster-related spectral profiles and peak positions show good agreement. This supports the assignment of the observed shifted features to Kr clusters formed in HNDs. After subtracting the gaseous Kr background, no significant atomic or dimer signals are observed in the spectra of HNDs. This systematic suppression of atomic and dimer species represents a key advantage over free expansion. However, because the subtraction procedures differ, the relative fractions of uncondensed atoms and clusters cannot be compared quantitatively from Fig.~\ref{fig:hatsui_comp}.

To investigate the influence of the initial HND size on the observed cluster shifts, complementary measurements were performed at three nozzle temperatures (13, 16, and 17~K) for each of two fixed effective Kr pickup pressures, approximately $1.34\times10^{-3}$ and $3.1\times10^{-3}$~mbar. Varying the nozzle temperature changes the initial HND size and, thereby, the amount of He available for dopant pickup and subsequent evaporation. The corresponding measurements and \texttt{HeDopant} simulations are presented in the Supplementary Material Fig.~S4. Over an initial HND size range from $\langle N_{\mathrm{He}}\rangle=7500$ to $64000$,
the mean Kr cluster size increases by a factor of six to seven relative to that obtained for the smallest HNDs, depending on the pickup pressure.
Despite the pronounced simulated cluster growth, for both pickup pressures the fitted cluster--gas energy shifts vary by no more than approximately $0.17$~eV over the investigated HND-size range.
Thus, the weak dependence of the shift cannot be explained by the formation of similarly sized Kr clusters; instead, it indicates that the binding-energy shift is only weakly sensitive to further cluster growth, consistent with the pickup-pressure dependence shown in Fig.~\ref{fig:Kr_shift_vs_doping}. This suggests that the local coordination and electronic screening responsible for the shift are already largely cluster-like even for the smallest clusters probed here.

Additional measurements performed over the photon-energy range 120--240 eV are presented in the Supplementary Material to verify that the observed cluster shifts are intrinsic and do not depend on the excitation energy. Within the experimental uncertainties, no significant dependence of the extracted shift on the photon energy was measured.

\subsection{Auger electron spectra}

To obtain complementary information on Kr aggregation, electron spectra covering the Kr $3d$ Auger-decay region were recorded under strong-doping conditions. Following direct ionization of the Kr $3d$ shell, the resulting $M_{4,5}$ core vacancy relaxes predominantly through local $MNN$ Auger decay: an electron from the outer $N$ shell fills the $3d$ vacancy, and the released energy ejects a second electron. The $MNN$ spectrum comprises several decay channels. In particular, the high-energy region between approximately $52$ and $59$~eV is associated mainly with the $M_{4,5}N_{2,3}N_{2,3}$ channel, which produces a local $4p^{-2}$ final state, whereas additional Auger and multielectron contributions occur at lower kinetic energies~\cite{Palaudoux2010,KuestnerWeteka2023,Hans2024}.

\begin{figure}[t]
\includegraphics[width=\columnwidth]{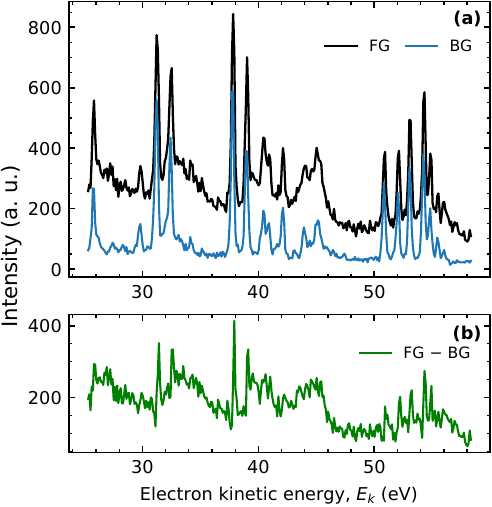}
\caption{
Kr Auger electron spectra recorded at $h\nu=160$~eV for Kr-doped HNDs at an effective Kr pickup pressure of $1.2\times10^{-3}$~mbar.
(a) Foreground (FG) and background (BG) spectra.
(b) Difference spectrum obtained by direct subtraction of BG from FG without applying an additional relative energy alignment. The narrow residual structures resulting from the incomplete cancellation of the
atomic Auger lines are discussed in the text.
}
\label{fig:Kr_Auger_difference}
\end{figure}

Figure~\ref{fig:Kr_Auger_difference} shows Auger electron spectra measured at $h\nu=160$~eV for a Kr pickup pressure of $1.2\times10^{-3}$~mbar. The foreground spectrum was recorded with the doped HND beam reaching the interaction region, whereas the background spectrum was acquired with the HND beam mechanically blocked. The spectra are overall similar in that both contain a dominant contribution from gas-phase Kr atoms present in the interaction region. 

Subtracting BG from FG reduces this stationary gas-phase contribution and reveals additional spectral contributions associated with the doped HND beam. No additional relative energy alignment was applied to the Auger spectra, since a rigid shift could correct small differences in peak position but not simultaneous differences in intensity and width. As a result, the sharp atomic Auger lines do not cancel out completely, leaving narrow and partly asymmetric residual structures in the difference spectrum that should not be interpreted as Kr-cluster Auger features.

From the \texttt{HeDopant} simulations we obtain an average pickup of $\langle N_{\mathrm{Kr}}\rangle\approx29$ Kr atoms per droplet, while the mean residual HND size decreases to $\langle N_{\mathrm{He,after}}\rangle\approx1300$ He atoms. The substantial HND depletion suggests that the cluster-associated Auger signal originates mostly from Kr aggregates formed in strongly depleted HNDs.

As shown in Fig.~\ref{fig:Kr_Auger_difference}(b), the FG$-$BG difference spectrum contains residual intensity extending from approximately $25$ to $58$~eV. Several relatively narrow and partly asymmetric residual structures, particularly in the $31{-}33$ and $38{-}40$~eV regions, are superimposed on a broad underlying intensity distribution. Weaker modulations are also present in the $25{-}28$ and $51{-}56$~eV regions.

Focusing on the broad underlying intensity rather than on the narrow residual lines, the cluster-associated Auger electron features can be
described approximately as an envelope of overlapping, broadened lines shifted overall by about $3$~eV toward higher electron kinetic energy relative to the
sharp atomic Auger lines. The overlapping groups of features have effective widths of the order of $2$~eV FWHM. These values are order-of-magnitude estimates from visual inspection of the spectra and do not result from a component-resolved fit. The shifted and broadened components are not equally apparent for every Auger line likely because of the irregular grouping of the atomic lines carrying variable intensities.

For reference, normal Auger spectra of considerably larger free Kr clusters exhibit surface- and bulk-site contributions shifted by approximately $2.3$ and $3.2$~eV, respectively, toward higher kinetic energy relative to the corresponding atomic lines~\cite{Peredkov2005}.
The broad components observed here in the
$M_{4,5}N_{2,3}N_{2,3}$ region are qualitatively compatible with an unresolved combination of such site-dependent cluster contributions, but the limited quality of the present spectrum does not allow their shifts to be determined separately. Site-specific Auger spectra have been isolated using electron--electron coincidence measurements on free Kr
clusters~\cite{KuestnerWeteka2023}; in the present non-coincidence measurement, contributions from different sites and from ionization of both Kr $3d$ spin--orbit components overlap~\cite{Hans2024}. The broad HND-dependent Auger contribution, together with the cluster-related photoelectron components and the pickup simulations, therefore provides complementary evidence for Kr aggregation inside the HNDs. Future improvements in the signal-to-noise ratio and in the suppression of the gas-phase atomic background should enable a transition-by-transition assignment of Auger electron spectra from dopant clusters formed in HNDs.

\section{Conclusion and Outlook}

In this work we have reported the first soft x-ray photoelectron spectra of dopant clusters formed in HNDs. Detailed simulations of the pickup process show that the strong-doping conditions used here to achieve sufficient signal-to-background ratios lead to substantial depletion of the HNDs by evaporation of He. Thus, the observed cluster spectra arise from compact dopant clusters formed in HNDs which are strongly depleted of He, approaching the limit of nearly bare dopant clusters at the highest doping levels. Kr and Ar core-level spectra show distinct gas-phase and cluster-related components, where the Kr cluster shifts are consistent with literature spectra for bare Kr clusters. The weak dependence of the Kr gas-to-cluster shift on the vapor pressure of dopants in the pickup cell, the initial HND size, and photon energy, indicates that the local electronic environment of the ionized atoms remains close to that of condensed Kr over the investigated range of simulated mean Kr cluster sizes, from $\langle N_{\mathrm{Kr}}\rangle\approx 10$ to 190. 

An important objective for future experiments is to distinguish the influence of the surrounding He shell on photoelectron spectra from other effects of dopant aggregation in view of assessing the potential of HNDs as matrices for XPS of embedded molecules and clusters. This would require photoelectron and Auger-electron spectra to be recorded at much lower doping levels with a mean number of picked up atoms around unity or below. Under otherwise identical conditions, and assuming that the dopant-derived electron yield scales approximately with the mean number of Kr atoms per droplet, reducing the mean pickup number from $\langle N_{\mathrm{Kr}}\rangle\approx29$ to e.\,g.  $0.1{-}0.3$ would lower the total HND-associated electron count rate by roughly two orders of magnitude. This estimate accounts only for the reduced number of dopants;
the detected single-atom signal may be further affected by electron scattering in the surrounding He shell.

The presented results establish HNDs as a route for preparing dopant clusters for soft-x-ray photoelectron and Auger-electron spectroscopy. The advantage of the HND technique is that multi-component clusters and aggregates can be formed of different types of atoms and molecules with a high degree of control with respect to the number and order of adding individual species; The latter can greatly differ in their physico-chemical properties such that e.~g. mixed clusters of refractory metals and fragile organic molecules can be assembled~\cite{haberfehlner2015formation,schiffmann2020helium,kollotzek2022helium}. Future experiments with improved background suppression, droplet-size selection, and electron-ion coincidence detection should allow more systematic studies of microsolvation~\cite{denifl2010ionization,fuchs2018microsolvation,ltaief2026tracking}, cluster-size effects~\cite{kollotzek2022efficient}, and ultrafast relaxation processes in doped HNDs~\cite{kautsch2013electronic,Mudrich2020}. In particular, extending this approach to near-edge x-ray absorption spectroscopy (XAS) would provide an element- and site-sensitive probe of the electronic structure of clusters and molecular aggregates formed in HNDs.

\section*{Acknowledgements}

We acknowledge SOLEIL for provision of synchrotron radiation facilities and we would like to thank the PLEIADES staff for assistance in preparing the experiment and using the beamline under proposal 20241684. We thank Dr.~B.~Zielinski for proofreading the manuscript and providing valuable suggestions for improvement.
We gratefully acknowledge financial support from the Deutsche Forschungsgemeinschaft (DFG, German Research Foundation) – SFB1319 – Projektnummer 328961117. N.S. and M.M. acknowledge support from the Novo Nordisk Foundation (grant no. NNF23OC0085401).
The research leading to these results has been supported by the COST Action CA21101 ``Confined Molecular Systems: From a New Generation of Materials to the Stars (COSY)''.

\bibliography{Bib}

\end{document}

\begin{center}

{\normalsize\bfseries Supplementary Information for:\par}
\vspace{0.5em}

{\large\bfseries
X-ray photoelectron spectroscopy of Ar and Kr clusters formed
in He nanodroplets\par}

\vspace{0.8em}

{\footnotesize
N. S. Blaj,$^{1}$ N. Scheel,$^{2}$ R. Rajni,$^{1}$
A. Ø. Lægdsmand,$^{1}$ J. Sadasivan,$^{3}$ S. De,$^{4}$\\
S. R. Krishnan,$^{5}$ J. Bozek,$^{6}$
A. R. Milosavljevi\'c,$^{6}$ and M. Mudrich$^{1,7,*}$\par}

\vspace{0.6em}

{\scriptsize\itshape
$^{1}$Institute of Physics, University of Kassel, 34132 Kassel, Germany\par
$^{2}$Department of Physics and Astronomy, Aarhus University,
8000 Aarhus C, Denmark\par
$^{3}$Department of Physics, Mahindra University, Hyderabad, India\par
$^{4}$Sorbonne Université, CNRS, Laboratoire de Chimie Physique
Matière et Rayonnement, UMR 7614, F-75005 Paris, France\par
$^{5}$Department of Physics and QuCenDiEM-group,
Indian Institute of Technology Madras, Chennai 600036, India\par
$^{6}$Synchrotron SOLEIL, St. Aubin, BP48,
91192 Gif sur Yvette Cedex, France\par
$^{7}$Center for Interdisciplinary Nanostructure Science and Technology
(CINSaT), University of Kassel, 34132 Kassel, Germany\par}

\end{center}

\section{Experimental details}
\subsection{Comparison of HeNDS calculations with literature data}

\begin{figure}[!h]
    \centering
    \includegraphics[width=0.65\columnwidth]{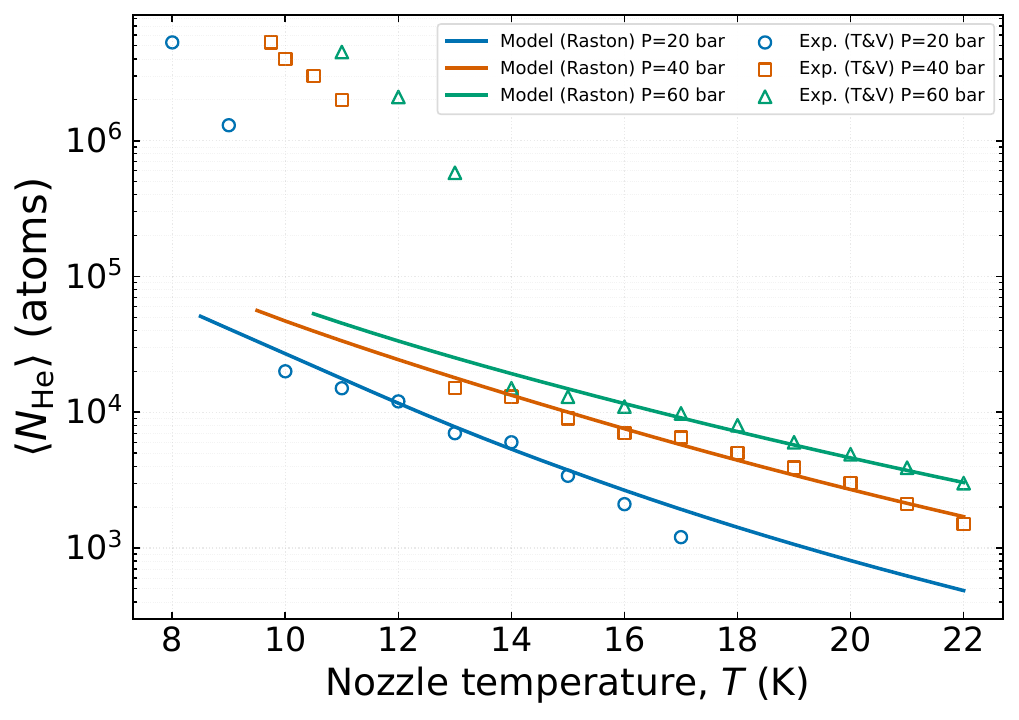}
    \caption{
Mean HND size $\langle N \rangle$ as a function of nozzle temperature $T$ for selected stagnation pressures.
Symbols denote values digitized from Toennies \& Vilesov~\protect\cite{Toennies2004}, while solid lines correspond to calculations using the HeNDS model by Raston~\protect\cite{raston2021hends}.
    }
    \label{fig:HeNDs_size_validation}
\end{figure}

\subsection{Electron spectrometer settings}

All photoelectron spectra were acquired with a Scienta hemispherical electron analyzer operated in swept acquisition mode.

The overall instrumental energy resolution was estimated from the monochromator bandwidth and the analyzer resolution according to

\begin{equation}
\Delta E_{\mathrm{tot}}
=
\sqrt{
\Delta E_{\mathrm{mono}}^{2}
+
\Delta E_{\mathrm{HEA}}^{2}
},
\end{equation}

where $\Delta E_{\mathrm{mono}}$ is determined by the photon energy and monochromator exit-slit setting, whereas $\Delta E_{\mathrm{HEA}}$ depends on the selected pass energy and analyzer entrance slit. For the acquisition conditions employed in this work, the estimated overall instrumental energy resolution ranged from approximately 0.29 to 0.54~eV. 

Most Kr $3d$ spectra were acquired with a pass energy of 100~eV using analyzer entrance slits between 600 and 700~µm. Depending on the experimental requirements, electron-energy step sizes between 0.05 and 0.1~eV were employed. Representative acquisition parameters include the high-statistics Kr $3d$ spectrum shown in Fig.~3, which was recorded using a step size of 0.05~eV, a dwell time of 200~ms, and 150 sweeps, and the representative Ar $2p$ spectrum, which was recorded using a step size of 0.1~eV, a dwell time of 500~ms, and 30 sweeps.

\subsection{Estimation of the effective pickup pressure}

The pressure measured with the capacitance gauge on the dopant-gas inlet line, denoted $P_{\mathrm{cell,gauge}}$, provides a direct experimental estimate of the pressure inside the pickup cell. However, this reading was not recorded for all photoelectron spectra, whereas the pressure in the surrounding doping chamber, denoted $P_{\mathrm{ch}}$, was available throughout the experiment. An empirical conversion was therefore established using measurements for which both pressure readings were available.

The background pressure in the doping chamber measured in the absence of Kr, denoted $P_{\mathrm{ch,BG}}$, was subtracted from $P_{\mathrm{ch}}$. Since the chamber gauge was gas dependent, the resulting pressure difference was corrected for its sensitivity to Kr,

\begin{equation}
\Delta P_{\mathrm{ch,Kr}}
=
f_{\mathrm{Kr}}
\left(
P_{\mathrm{ch}}-P_{\mathrm{ch,BG}}
\right),
\end{equation}

where $f_{\mathrm{Kr}}=0.5$ and
$P_{\mathrm{ch,BG}}=9.57\times10^{-7}$~mbar. The effective pickup pressure was then obtained from

\begin{equation}
P_{\mathrm{cell}}
=
a\,\Delta P_{\mathrm{ch,Kr}}.
\end{equation}

The proportionality factor $a$ was determined by a least-squares fit constrained to pass through the origin using the paired measurements of $P_{\mathrm{cell,gauge}}$ and $\Delta P_{\mathrm{ch,Kr}}$. The fit yielded $a=8.54\times10^{2}$. This empirical relation was applied consistently to all measurements to obtain a common scale of effective pickup pressures, including spectra for which no direct capacitance-gauge reading was available. The resulting calibrated pressures were also used as input parameters for the \texttt{HeDopant} simulations.

\section{Energy calibration and data processing}

\subsection{Kinetic-energy calibration}

The kinetic-energy scale of the hemispherical electron analyzer was calibrated using gas-phase Kr 3d spectra recorded at photon energies between 120 and 240 eV. For each photon energy, the expected positions of the atomic $3d_{5/2}$ and $3d_{3/2}$ photoelectron lines were calculated from the corresponding literature binding energies according to
\begin{equation}
E_{K,\mathrm{ref}}(h\nu) = h\nu-E_{B,\mathrm{ref}},
\end{equation}
using literature binding energies of
$E_{B,\mathrm{ref}}(3d_{5/2})=93.8$~eV and
$E_{B,\mathrm{ref}}(3d_{3/2})=95.0$~eV \cite{Vaughan1986}.

The photon-energy-dependent offset of the analyzer scale was defined as
\begin{equation}
\Delta E_{\mathrm{cal}}(h\nu) = E_{K,\mathrm{ref}}(h\nu) - E_{K,\mathrm{meas}}(h\nu).
\end{equation}
The calibration offsets obtained from the two spin-orbit components were consistent within the experimental uncertainties. A single photon-energy-dependent calibration function was therefore used for all spectra
\begin{equation}
\Delta E_{\mathrm{cal}}(h\nu) = 0.03587\,h\nu\left[\mathrm{eV}\right]-2.336\,\, \mathrm{eV}.
\end{equation}
The corrected kinetic energies were then calculated as
\begin{equation}
E_{K,\mathrm{corr}} = E_{K,\mathrm{meas}} + \Delta E_{\mathrm{cal}}(h\nu),
\end{equation}
and the corresponding binding energies as
\begin{equation}
E_{B,\mathrm{corr}} = h\nu-E_{K,\mathrm{corr}}.
\end{equation}
The same photon-energy-dependent correction was applied to all foreground and background spectra recorded at the corresponding photon energies.

\section{Line-shape model and fitting procedure}

\subsection{Model definition}

Each spectrum was modeled as a sum of $N$ asymmetric spectral components and a quadratic background,

\begin{equation}
I(E)
=
\sum_{i=1}^{N} S_i(E)
+
aE^2+bE+c.
\end{equation}

Each component was described using the location--scale form of the skew-normal density introduced by Azzalini~\cite{Azzalini1985}, with the normal cumulative distribution function expressed in terms of the error function

\begin{equation}
S(E)
=
A
\exp\!\left[
-\frac{1}{2}
\left(
\frac{E-\mu}{\sigma}
\right)^2
\right]
\left[
1+
\operatorname{erf}\!\left(
\frac{\alpha}{\sqrt{2}}
\frac{E-\mu}{\sigma}
\right)
\right].
\label{eq:skew_normal_standard}
\end{equation}

Here, $A$ is a normalization amplitude, $\mu$ is a location parameter, $\sigma>0$ is a scale parameter, and $\alpha$ is a dimensionless shape parameter controlling the asymmetry. For $\alpha=0$, Eq.~\eqref{eq:skew_normal_standard} reduces to a symmetric Gaussian, for which $\mu$ and $\sigma$ coincide with the mean and standard deviation. For $\alpha\neq0$, however, $\mu$ is not the mean of the distribution and $\sigma$ is not its standard deviation.

For the quantitative analysis, the conventional parameters $A$, $\mu$, and $\sigma$ were reparametrized in terms of the integrated area $Q$, mean electron energy $\bar{E}$, and standard deviation $s$, while retaining $\alpha$ as the shape parameter. This one-to-one reparametrization does not change the underlying line shape or the set of profiles represented by the model; an equivalent fit could therefore be performed using $A$, $\mu$, $\sigma$, and $\alpha$. It was adopted to make the physically interpretable spectral quantities $Q$, $\bar{E}$, and $s$ directly accessible during the fit and to allow constraints, such as fixed area ratios and mean-energy separations, to be imposed directly on these quantities.

These quantities are defined by

\begin{equation}
Q
=
\int_{-\infty}^{\infty}
S(E)\,\mathrm{d}E,
\end{equation}

\begin{equation}
\bar{E}
=
\frac{1}{Q}
\int_{-\infty}^{\infty}
E\,S(E)\,\mathrm{d}E,
\end{equation}

and

\begin{equation}
s^2
=
\frac{1}{Q}
\int_{-\infty}^{\infty}
(E-\bar{E})^2 S(E)\,\mathrm{d}E.
\end{equation}

For the conventional skew-normal profile, these quantities are related to $A$, $\mu$, and $\sigma$ by

\begin{equation}
Q
=
A\sigma\sqrt{2\pi},
\end{equation}

\begin{equation}
\bar{E}
=
\mu
+
\sigma\delta\sqrt{\frac{2}{\pi}},
\end{equation}

and

\begin{equation}
s^2
=
\sigma^2 k,
\end{equation}

where

\begin{equation}
\delta
=
\frac{\alpha}{\sqrt{1+\alpha^2}}
\end{equation}

is a bounded, dimensionless transformation of the shape parameter, and

\begin{equation}
k
=
1-\frac{2\delta^2}{\pi}
\end{equation}

is the corresponding variance-correction factor. Neither $\delta$ nor $k$ is an independent fit parameter.

Solving these relations for the conventional parameters gives

\begin{equation}
\sigma
=
\frac{s}{\sqrt{k}},
\end{equation}

\begin{equation}
\mu
=
\bar{E}
-
\frac{s\delta}{\sqrt{k}}
\sqrt{\frac{2}{\pi}},
\end{equation}

and

\begin{equation}
A
=
\frac{Q\sqrt{k}}{s\sqrt{2\pi}}.
\end{equation}

The spectral component used in the fitting procedure is therefore

\begin{equation}
\begin{split}
S(E;Q,\bar{E},s,\alpha)
={}&
\frac{Q\sqrt{k}}{s\sqrt{2\pi}}
\exp\!\left\{
-\frac{1}{2}
\left[
\frac{\sqrt{k}}{s}(E-\bar{E})
+
\delta\sqrt{\frac{2}{\pi}}
\right]^2
\right\}
\\
&\times
\left\{
1+
\operatorname{erf}\!\left[
\frac{\alpha}{\sqrt{2}}
\left(
\frac{\sqrt{k}}{s}(E-\bar{E})
+
\delta\sqrt{\frac{2}{\pi}}
\right)
\right]
\right\}.
\end{split}
\label{eq:skew_normal_qmsa}
\end{equation}

Thus, the independent parameters optimized in the fitting procedure are $Q$, $\bar{E}$, $s$, and $\alpha$. By construction, Eq.~\eqref{eq:skew_normal_qmsa} has integrated area $Q$, mean electron energy $\bar{E}$, and standard deviation $s$. For $\alpha=0$, it reduces to a Gaussian centered at $\bar{E}$ with standard deviation $s$.

\subsection{Constraints and shared parameters}

To stabilize the multi-component decomposition, selected pairs of components were constrained using fixed intensity ratios and fixed energy separations.
For dependent components, the constraints were implemented as

\begin{equation}
Q_{\mathrm{dep}} = f\, Q_{\mathrm{ref}},
\end{equation}

\begin{equation}
\bar{E}_{\mathrm{dep}} = \bar{E}_{\mathrm{ref}} + \Delta E,
\end{equation}

where $f$ and $\Delta E$ were fixed parameters chosen based on physically motivated considerations (e.g., spin–orbit splitting). At lower photon energies, additional spectral components were required in the FG spectra due to overlapping Auger-related and satellite contributions, which complicated the decomposition of the cluster-related photoelectron features.

\section{Simulations using the H\lowercase{e}D\lowercase{opant} package}

The pickup and aggregation dynamics inside HNDs were simulated using the \texttt{HeDopant} code developed by Sishodia and De~\cite{HeDopant}.
The model describes the evolution of dopant-cluster populations during the passage of a HND through a pickup cell, explicitly accounting for pickup statistics, energy deposition, He evaporation, and the resulting modification of the droplet size during the pickup process. The pickup of dopant atoms by HNDs is commonly described using Poisson statistics with a constant pickup rate~\cite{Lewerenz1993,Lewerenz1995}. In \texttt{HeDopant}, however, the pickup rate is state dependent because the size and velocity of the HND are updated after each pickup event and the associated evaporation of He atoms. The HND was approximated as a spherical droplet with a uniform, bulk-like He number density $\rho_{\mathrm{He}}$, consistent with the liquid-drop description of HNDs~\cite{Toennies2004}. Its radius is therefore given by

\begin{equation}
R(N_{\mathrm{He}})
=
\left(
\frac{3N_{\mathrm{He}}}
{4\pi\rho_{\mathrm{He}}}
\right)^{1/3},
\end{equation}

where $N_{\mathrm{He}}$ is the instantaneous number of He atoms in the
droplet. The geometrical pickup cross section used in the simulations
was taken as the projected area of the HND,

\begin{equation}
\sigma_{\mathrm{pickup}}(N_{\mathrm{He}})
=
\pi R^2(N_{\mathrm{He}})
=
\pi
\left(
\frac{3N_{\mathrm{He}}}
{4\pi\rho_{\mathrm{He}}}
\right)^{2/3}.
\label{eq:pickup_cross_section}
\end{equation}

Here, $\pi R^2$ represents the projected geometrical collision area of the HND. The total surface area, $4\pi R^2$, is not the appropriate pickup cross section. This geometrical cross section is independent of the relative magnitudes of the droplet and dopant velocities, whose effect on the collision rate is included separately through the relative-velocity factor discussed below. Consequently, $\sigma_{\mathrm{pickup}}\propto N_{\mathrm{He}}^{2/3}$ and decreases as He atoms evaporate during sequential dopant pickup. This produces a history-dependent pickup rate that is absent from the conventional Poisson model.

\subsection{Relative-velocity treatment}

In the original \texttt{HeDopant} implementation, the dopant velocity entering the relative-velocity factor is provided through the dopant species configuration file. The pickup coefficient contains the factor

\begin{equation}
\frac{v_{\mathrm{rel}}}{v_{\mathrm{drop}}}
=
\sqrt{1+\frac{v_{\mathrm{dop}}^2}{v_{\mathrm{drop}}^2}},
\end{equation}

which is equivalent to

\begin{equation}
v_{\mathrm{rel}}
=
\sqrt{
v_{\mathrm{drop}}^2+
v_{\mathrm{dop}}^2
}.
\end{equation}

In the present calculations, the dopant velocity was chosen as the RMS thermal velocity,

\begin{equation}
v_{\mathrm{dop}}
=
\sqrt{\frac{3k_{\mathrm B}T}{m_X}},
\end{equation}

where $T$ is the dopant-gas temperature and $m_X$ is the dopant mass. This choice makes the velocity input consistent with the thermal Maxwell-Boltzmann motion of the dopant gas, while retaining the original relative-velocity structure used in \texttt{HeDopant}.

\subsection{Simulation parameters}

Unless stated otherwise, simulations were performed for initial HND sizes of
$\langle N_{\mathrm{He}}\rangle = 9678$,
corresponding approximately to the experimental nozzle conditions at $T=16$~K. The Kr-gas temperature was assumed to be $T=300$~K.
Pickup outside the dedicated pickup cell was neglected in the simulations. 
Using the effective interaction lengths and the experimentally determined pressure-conversion factor, the contribution from the surrounding effusive rare-gas background is estimated to be only a few percent of the pickup occurring in the dedicated doping cell.

\subsection{Size-dependent interaction energy}

The dopant--dopant interaction energy was modified using a size-dependent cohesive-energy model motivated by literature data for rare-gas clusters~\cite{Schwerdtfeger2006}.

\begin{equation}
\Delta E(N) = aN^{-1/3} + b,
\end{equation}

where $a$ is the slope describing the finite-size dependence of the cohesive energy, while $b=\lim_{N\rightarrow\infty}\Delta E(N)$ is the bulk cohesive-energy limit. For each dopant species, $b$ was fixed
to the literature bulk value, whereas $a$ was obtained by fitting the digitized size-dependent data from Ref.~\cite{Schwerdtfeger2006}. This modification allows the model to reproduce both the correct dimer binding energy and the asymptotic bulk cohesive-energy limit.
Together, these modifications provide a more realistic description of the pickup dynamics and He depletion occurring under strong-doping conditions.

\section{Order-of-magnitude estimate of the emitted photoelectron rate}

To assess the consistency between the expected and experimentally measured electron detection rates, an order-of-magnitude estimate of the rate of emitted photoelectrons was performed. Assuming that each Kr atom embedded in a He nanodroplet contributes independently to the photoelectron signal, the emitted photoelectron rate can be approximated by

\begin{equation}
R_{\mathrm{emit}} = n_{\mathrm{HND}} \,\langle N_{\mathrm{Kr}}\rangle \,L\,\Phi\,\sigma ,
\end{equation}

where $n_{\mathrm{HND}}$ is the HND number density in the interaction region, $\langle N_{\mathrm{Kr}}\rangle$ is the simulated mean number of Kr atoms per HND, $L$ is the effective interaction length, $\Phi$ is the incident photon flux, and $\sigma_{\mathrm{Kr}\,3d}$ is the Kr $3d$ photoionization cross section.

For representative strong-doping conditions at $h\nu=160$~eV, the
values used in this estimate were
$n_{\mathrm{HND}}\approx5\times10^{7}$~cm$^{-3}$,
$\langle N_{\mathrm{Kr}}\rangle\approx29$,
$L=8$~mm $=0.8$~cm,
$\Phi\approx10^{13}$~photons/s, and
$\sigma_{\mathrm{Kr}\,3d}\approx5.4$~Mb
$=5.4\times10^{-18}$~cm$^2$.
The value of $\langle N_{\mathrm{Kr}}\rangle$ was obtained from the corresponding \texttt{HeDopant} simulation, whereas
$n_{\mathrm{HND}}$ was estimated from the source geometry and beam divergence. The photon flux represents an order-of-magnitude reference value based on the beamline calibration rather than an absolute measurement under the exact experimental settings. Substitution of these values gives

\begin{equation}
\begin{aligned}
R_{\mathrm{emit}}
&\approx
\left(5\times10^{7}\,\mathrm{cm}^{-3}\right)
(29)
(0.8\,\mathrm{cm})
\left(10^{13}\,\mathrm{s}^{-1}\right)
\left(5.4\times10^{-18}\,\mathrm{cm}^{2}\right)
\\
&\approx
6\times10^{4}\ \mathrm{electrons / s},
\end{aligned}
\end{equation}

i.e., an emitted photoelectron rate on the order of $10^{5}$~electrons/s before analyzer acceptance, transmission, and detector losses are taken into account.

The detected count rate can be expressed as

\begin{equation}
R_{\mathrm{det}}(E)
=
R_{\mathrm{emit}}\,G(E),
\end{equation}

where the overall detection factor is written as

\begin{equation}
G(E)
=
G_{\mathrm{HEA}}(E)\,
G_{\mathrm{MCP}}(E).
\end{equation}

Here, $G_{\mathrm{HEA}}(E)$ includes both the geometrical acceptance and the energy-dependent transmission of the hemispherical electron analyzer, while $G_{\mathrm{MCP}}(E)$ denotes the MCP detection efficiency.

The maximum background-subtracted experimental count rate at the Kr $3d$ peak was approximately
$R_{\mathrm{det,max}}\approx1.6\times10^{2}$~counts/s.
Comparison with the estimated emitted rate, $R_{\mathrm{emit}}\approx6\times10^{4}$~electrons/s, gives the rough peak effective detection factor

\begin{equation}
G_{\mathrm{eff}}^{\mathrm{peak}}
\approx
\frac{R_{\mathrm{det,max}}}{R_{\mathrm{emit}}}
\approx
\frac{1.6\times10^{2}}{6\times10^{4}}
\approx
3\times10^{-3}.
\end{equation}

Thus, the experimentally inferred overall detection factor is of the order of $10^{-3}$. This is qualitatively consistent with the combined losses expected from the finite acceptance and transmission of the HEA and the non-unity MCP detection efficiency. The inferred value should nevertheless be regarded as an order-of-magnitude estimate because $R_{\mathrm{emit}}$ represents the total emitted Kr $3d$ photoelectron rate, whereas the experimental value refers to the maximum count rate at the spectral peak rather than to an energy-integrated detection rate.

\clearpage

\section{Supplementary figures}

\subsection{Photon-energy dependence of the cluster shifts}

\begin{figure}[h]
\centering
\includegraphics[width=0.55\columnwidth]{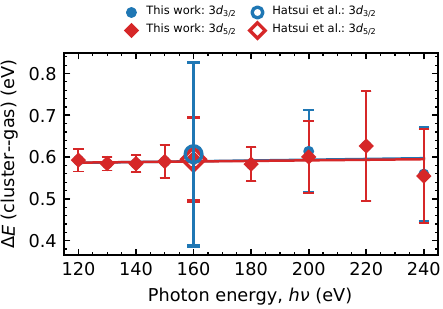}
\caption{
Cluster--gas energy shift $\Delta E$ of the Kr $3d_{3/2}$ and $3d_{5/2}$ photoelectron lines as a function of photon energy $h\nu$.
Filled symbols represent the present measurements, whereas open symbols denote digitized literature values for pure Kr clusters formed in a free expansion~\protect\cite{Hatsui}.
Solid lines correspond to linear fits; error bars denote $1\sigma$ fitting uncertainties.
}
\label{fig:Kr_shift_vs_hv}
\end{figure}

Figure~\ref{fig:Kr_shift_vs_hv} summarizes the photon-energy dependence of the extracted cluster--gas shifts for the Kr $3d$ photoelectron lines.
The measurements were performed at an effective Kr pickup pressure of approximately $1.3\times10^{-3}$~mbar. According to the \texttt{HeDopant} simulations, these conditions correspond to an average pickup of approximately $\langle N_{\mathrm{Kr}}\rangle \approx 29$ Kr atoms per droplet and a mean residual HND size after pickup of approximately $\langle N_{\mathrm{He,after}}\rangle \approx 1.4\times10^{3}$ He atoms.

Within the experimental uncertainties, the extracted cluster--gas energy shifts remain approximately constant over the investigated photon-energy range and are consistent with previous measurements of pure Kr clusters formed in a free expansion~\cite{Hatsui}.

\subsection{Influence of the initial HND size distribution}

\begin{figure*}[h]
\centering
\includegraphics[width=0.92\textwidth]{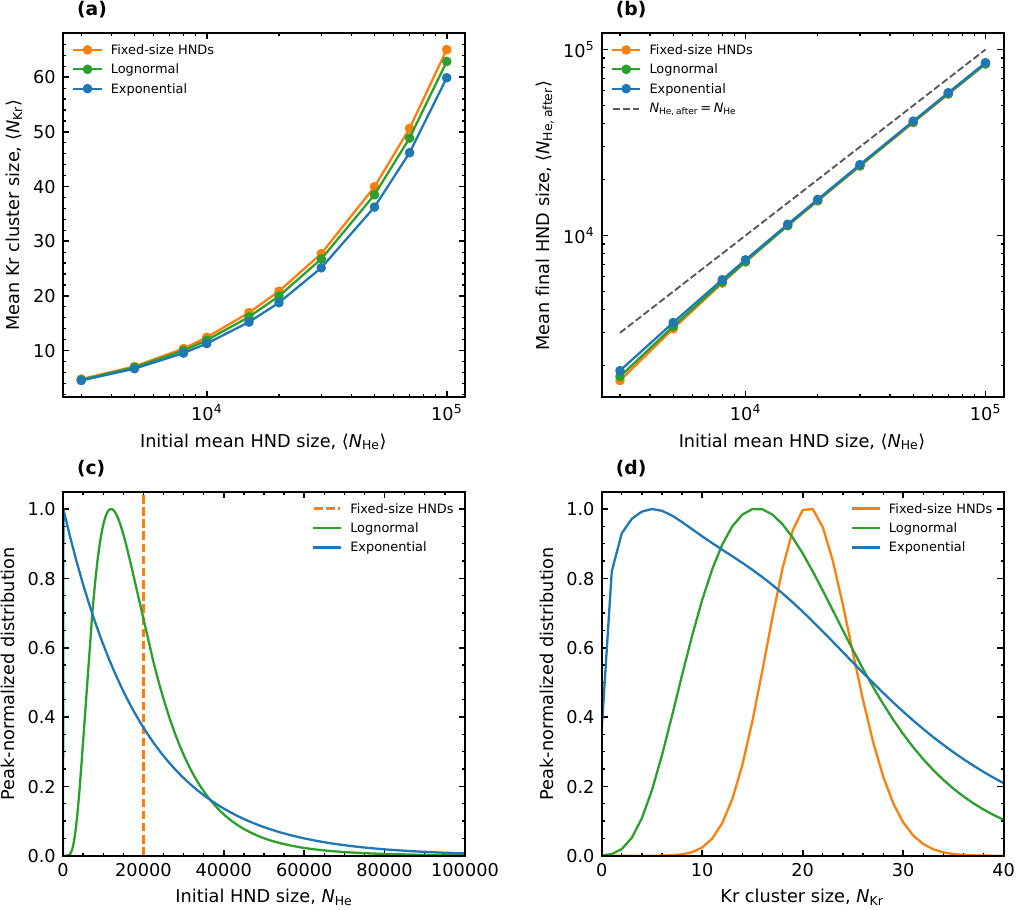}
\caption{
Influence of the assumed initial HND size distribution on the
\texttt{HeDopant} results obtained at a fixed effective Kr pickup
pressure of $P_{\mathrm{cell}}=3.0\times10^{-4}$~mbar.
(a) Simulated mean Kr cluster size, $\langle N_{\mathrm{Kr}}\rangle$, as a function of the initial mean HND size for fixed-size, lognormal, and exponential HND size distributions.
(b) Simulated mean HND size after pickup and evaporative cooling, $\langle N_{\mathrm{He,after}}\rangle$, for the same simulations. The dashed line indicates $\langle N_{\mathrm{He,after}}\rangle=\langle N_{\mathrm{He}}\rangle$, corresponding to the absence of evaporation.
(c) Initial HND size distributions assumed for simulations with $\langle N_{\mathrm{He}}\rangle=20000$. The vertical dashed line indicates the corresponding fixed-size HND simulation.
(d) Resulting Kr cluster-size distributions for $\langle N_{\mathrm{He}}\rangle=20000$.
For clarity, the continuous distributions in panel (c) and the Kr cluster-size distributions in panel (d) were independently normalized to their respective maximum values. In panel (d), the horizontal axis is limited to $N_{\mathrm{Kr}}\leq40$ for clarity.
}
\label{fig:HND_distribution_effects}
\end{figure*}

To evaluate the influence of the assumed initial HND size distribution on the pickup simulations, \texttt{HeDopant}
calculations were performed at a fixed effective Kr pickup pressure of $P_{\mathrm{cell}}=3.0\times10^{-4}$~mbar using
fixed-size, lognormal, and exponential HND size distributions.
As shown in Fig.~\ref{fig:HND_distribution_effects}(a,b), the simulated mean Kr cluster size and mean residual HND size depend only weakly on the assumed distribution shape over the investigated range of initial mean HND sizes.

The differences between the distributions are examined in more detail for $\langle N_{\mathrm{He}}\rangle=20000$ in Fig.~\ref{fig:HND_distribution_effects}(c,d). Although the three initial distributions have the same mean HND size, their shapes differ markedly. The fixed-size model assigns the same initial size to every HND, whereas the lognormal distribution contains droplets both smaller and larger than the mean and the exponential distribution is dominated by small droplets while retaining a long tail toward large sizes. Because the curves in panels (c) and (d) were independently normalized to their maxima, these panels compare their shapes, widths, and peak positions rather than their absolute populations.

These differences are retained in the resulting Kr cluster-size distributions shown in Fig.~\ref{fig:HND_distribution_effects}(d). The fixed-size model produces a relatively narrow distribution centered near $N_{\mathrm{Kr}}\approx20$, whereas the lognormal model gives a broader distribution with a maximum near $N_{\mathrm{Kr}}\approx16$. The exponential HND distribution produces the broadest and most asymmetric Kr distribution, with a maximum near $N_{\mathrm{Kr}}\approx5$ and a pronounced tail toward larger cluster sizes that remains significant at the upper limit of the displayed range.

Although small Kr clusters are most probable for the exponential HND distribution, the less probable large clusters in its extended high-size tail shift the mean toward larger values. Consequently, the mean Kr cluster sizes in panel (a) remain relatively similar even though the peak positions and shapes of the distributions in panel (d) differ markedly. Thus, under the present simulation conditions, conclusions based on the mean Kr cluster size and mean residual HND size are relatively insensitive to the assumed initial HND size distribution, whereas the predicted distribution of individual Kr cluster sizes is strongly model dependent.

\FloatBarrier 

\subsection{Nozzle-temperature dependence of the Kr spectra}

\begin{figure*}[h]
\centering
\includegraphics[width=0.92\textwidth]{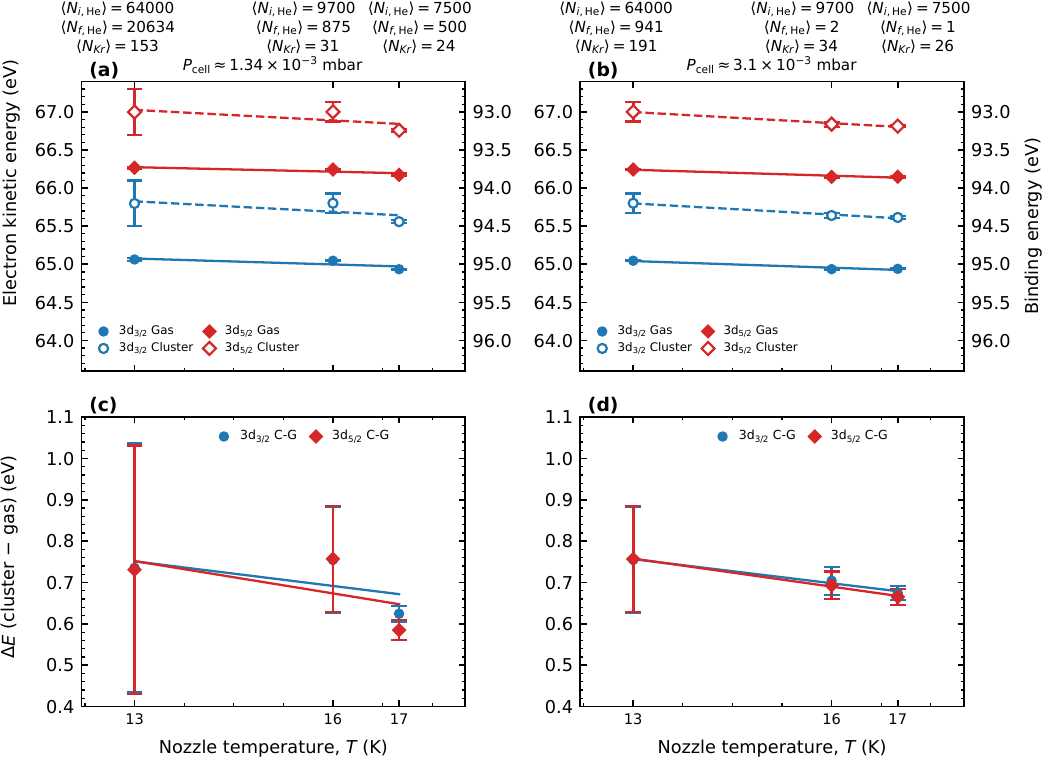}
\caption{
Dependence of the fitted Kr $3d$ peak positions and cluster--gas energy shifts on the initial HND size for two different Kr pickup conditions. All spectra were recorded at a photon energy of $h\nu = 160$~eV.
Panels (a) and (c) correspond to an effective Kr pickup  pressure of approximately $1.34\times10^{-3}$~mbar, whereas panels (b) and (d) correspond to approximately $3.1\times10^{-3}$~mbar.
The upper panels show the electron kinetic energies of the gas-phase and cluster-related $3d_{3/2}$ and $3d_{5/2}$ components, while the lower panels show the corresponding cluster--gas energy shifts extracted from the spectral fits.
The values above the upper panels indicate the estimated initial mean HND size, the simulated mean residual HND size after pickup and evaporation, and the corresponding mean Kr cluster size.
Solid lines represent linear fits to the data. Error bars denote $1\sigma$ fitting uncertainties.
}
\label{fig:Kr_nozzle_dependence}
\end{figure*}

Figure~\ref{fig:Kr_nozzle_dependence} summarizes the evolution of the fitted Kr $3d$ peak positions and cluster--gas energy shifts as a function of the estimated initial HND size. Two sets of measurements are shown, corresponding to effective Kr pickup-cell pressures of approximately $1.34\times10^{-3}$ and $3.1\times10^{-3}$~mbar. All spectra were recorded at a photon energy of $h\nu=160$~eV.
According to the \texttt{HeDopant} simulations, decreasing the nozzle temperature increases the initial HND size, allowing larger Kr clusters to form before pickup terminates. As the nozzle temperature is decreased from 17 to 13~K, the simulated mean Kr cluster size increases from approximately 24 to 153 atoms at the lower pickup pressure and from approximately 26 to 191 atoms at the higher pickup pressure. The corresponding simulated mean residual HND sizes after pickup and evaporative cooling are indicated together with the mean Kr cluster sizes in the figure. For both pickup conditions, the extracted cluster--gas energy shifts exhibit only modest changes despite these substantial variations in the simulated Kr cluster size and residual He content.

\clearpage

\section{Supplementary references}

{\footnotesize
\noindent $^{*}$mudrich@uni-kassel.de\par
}

\vspace{0.5em}

\bibliography{Bib}